\documentclass[aps,prb,floatfix,twocolumn,showpacs,superscriptaddress]{revtex4-2}  
\usepackage{graphicx}  
\usepackage{amsmath,amssymb}

\usepackage{physics}

\usepackage{flushend}
\usepackage[colorlinks = true,
            linkcolor = blue,
            urlcolor  = blue,
            citecolor = blue,
            anchorcolor = blue]{hyperref}
\usepackage{array}
\usepackage{float}
\usepackage{natbib}
\usepackage{booktabs}
\mathchardef\Re="023C
\mathchardef\Im="023D

\usepackage{color}
\usepackage{times}

\begin{document}

\title{Refraction-induced transverse charge transport}

\author{Ronika Sarkar} \email[]{ronikasarkar@iisc.ac.in} \affiliation{Department of Physics, Indian Institute of Science, Bangalore 560012, India}
\affiliation{Solid State and Structural Chemistry Unit, Indian Institute of Science, Bangalore 560012, India}
\author{Arka Bandyopadhyay} \email[]{arka.bandyopadhyay@uni-wuerzburg.de} \affiliation{Institute for Theoretical Physics and Astrophysics, University of Würzburg, 97074 Würzburg, Germany}
\author{Awadhesh Narayan} \email[]{awadhesh@iisc.ac.in} \affiliation{Solid State and Structural Chemistry Unit, Indian Institute of Science, Bangalore 560012, India}
\author{Diptiman Sen} \email[]{diptiman@iisc.ac.in} \affiliation{Centre for High Energy Physics, Indian Institute of Science, Bangalore 560012, India}

\vskip 0.25cm
\begin{abstract}
We introduce a new mechanism that produces a Hall-like transverse response in time-reversal-invariant materials, driven entirely by geometric effects. Specifically, we demonstrate that a tilted potential interface causes electron wave packets to undergo a refraction-like deflection upon transmission through the barrier, leading to a finite transverse current and a corresponding Hall-like conductance. Our analytical framework captures the essential features of this refraction-driven charge transport, and the resulting transverse conductance profile is corroborated by numerical simulations across different lattice models and device geometries. We further visualize our predicted effect through real-time wave packet dynamics, which reveals its purely geometric origin and the robustness of the transverse response. These findings establish a fundamentally distinct class of Hall-like transport phenomena in mesoscopic systems that preserve time-reversal symmetry.
\end{abstract}

\maketitle

\date{\today}

\section{Introduction}
Transport phenomena at interfaces and the quantum
Hall effect have long been central to our understanding of condensed matter systems and topological physics.
Examples of interfaces include a junction between the
surfaces of two topological insulators~\cite{takahashi2011,biswas2011,deb2012}
and a junction between topological insulators and a
superconductor~\cite{soori2013}.
Over the years, a variety of Hall effects have been identified, each revealing unique insights into the interplay between electron dynamics, external fields and material properties. Starting with the classical Hall effect~\cite{hall1879new}, which describes the transverse voltage generated under an applied magnetic field, numerous related phenomena have been discovered. These include the anomalous Hall effect~\cite{nagaosa2010anomalous,jungwirth2002anomalous,nakatsuji2015large}, where intrinsic spin-orbit coupling gives rise to a transverse charge current even without a magnetic field, and the spin Hall effect~\cite{sinova2015spin,hirsch1999spin,hoffmann2013spin,valenzuela2006direct,kato2004observation}, which generates a transverse spin current in the absence of a charge current. In the quantum regime, extensive studies have focused on the integer and fractional quantum Hall effects~\cite{klitzing1980new,prange1987quantum,stone1992quantum,stormer1999fractional,moore1991nonabelions,bolotin2009observation}, the quantum spin Hall effect~\cite{bernevig2006quantum,kane2005quantum,qi2010quantum,bernevig2006quantum}, and the quantum anomalous Hall effect~\cite{haldane1988model,chang2013experimental,liu2016quantum}. More recently, the thermal Hall effect~\cite{katsura2010theory,banerjee2018observation,kasahara2018majorana} has attracted attention, linking temperature gradients to transverse currents, while additionally, non-linear Hall effects~\cite{sodemann2015quantum,du2021nonlinear,ortix2021nonlinear,bandyopadhyay2024non} have emerged as a new class of phenomena where the intrinsic Berry curvature dipole of a system leads to a second-order transverse response to external electric fields. Furthermore, the Magnus Hall effect~\cite{papaj2019magnus,mandal2020magnus,costa2018gravitational} describes the generation of a linear transverse response in systems with an in-built electric field under inversion-symmetry breaking.

\begin{figure}[b]
    \centering
    \includegraphics[width=0.48\textwidth]{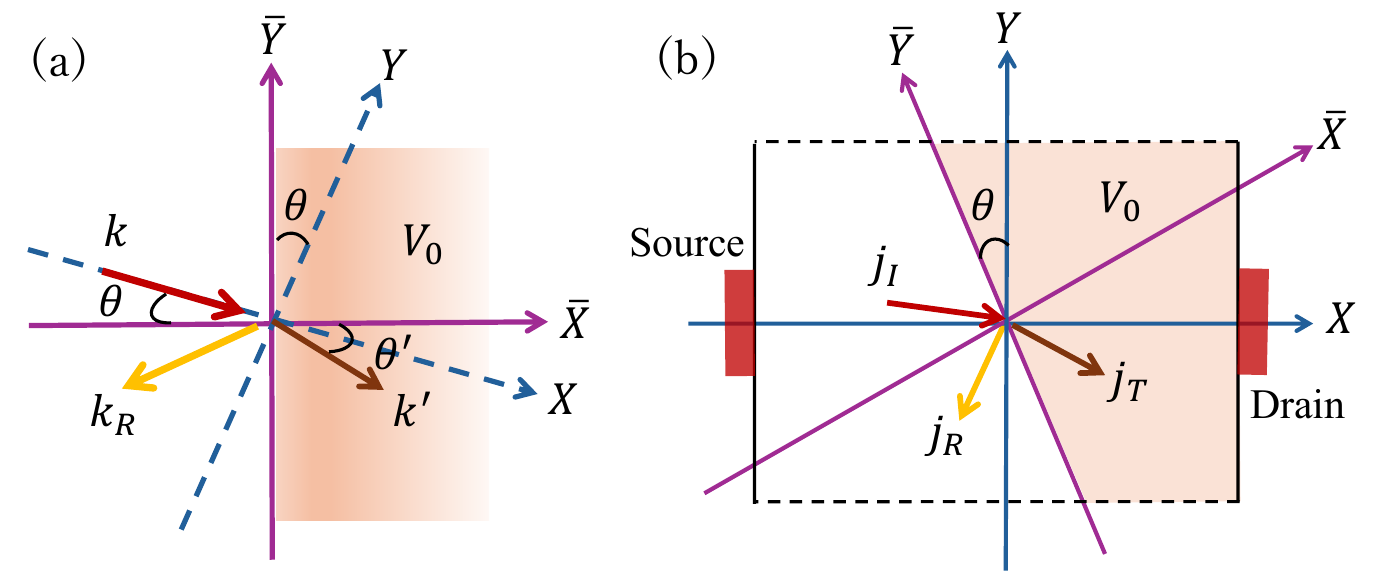}
    \caption{\label{Fig:Schematic} \textbf{Snell's law for electrons and schematic of the refraction setup.} (a) An electron with incident wavevector with magnitude $k$ and angle $\theta$ impinges on a potential step aligned with the $(\bar{X},\bar{Y})$ axes. Upon transmission, its wavevector changes to $k'$ and the refracted trajectory emerges at an angle $\theta'$, satisfying the conservation law $k \sin\theta = k' \sin\theta'$. (b) The device consists of a rectangular slab connected to a source and a drain along the $X-$axis and extending infinitely along $Y$. A potential barrier is engineered at an angle $\theta$ relative to the perpendicular axis ($Y$). The rotated coordinates ($\bar{X},\bar{Y}$) follow the orientation of the barrier. The incident current density from the source, $j_I$, represents a single illustrative angle of incidence. The corresponding reflected and transmitted components are indicated by $j_R$ and $j_T$, respectively, analogous to reflection and refraction in an optical setting.} 
\end{figure}

Although a vast body of research on Hall effects focuses on systems subject to magnetic fields, the possibility of generating transverse currents without breaking time-reversal symmetry remains relatively less explored. In this work, we investigate a geometric mechanism that induces a Hall-like response without using external magnetic or electric fields, arising from electron scattering at a tilted potential step. Using both analytical and numerical analyses under a small applied bias, we derive an expression for the differential transverse conductance and show that a Hall-like current emerges due to angular refraction effects, reminiscent of optical refraction.\\

We first analyze a single-electron case, where an electron refracts upon encountering a potential barrier at a fixed incidence angle, to compute the resulting transverse current. We then extend this to a biased device with a tilted potential interface, where electrons incident at arbitrary angles yield a transverse conductance dependent on the tilt angle and the potential strength. To further elucidate the dynamics, we simulate the time evolution of an electron wave packet in this geometry, visually illustrating the refraction-induced Hall-like response. Finally, we validate our theory through numerical simulations on square and hexagonal lattices, confirming the feasibility of the proposed effect in realistic experimental systems.\\

\section{Electron refraction at a tilted potential interface}
We begin by considering an electron with a wavevector of magnitude $k$ incident at an angle $\theta$ on a potential step aligned with the Cartesian axes $(\bar{X},\bar{Y})$. Upon transmission into the region of constant potential $V_0$, the electron experiences a refraction process directly analogous to Snell’s law in optics [see Fig.~\ref{Fig:Schematic}(a)]. In this region, the wavevector magnitude changes to $k'$, and the transmitted trajectory forms an angle $\theta'$ with the interface normal. Conservation of the parallel component of the wavevector leads to the relation, $k \sin\theta = k' \sin\theta'$.

In the context of our system depicted in Fig.~\ref{Fig:Schematic}(b), this configuration corresponds to an electron incident parallel to the sample from the source side, subsequently encountering the potential barrier at an angle $\theta$. The electron dynamics is governed by the Schr\"odinger equation expressed in the $(\bar{X},\bar{Y})$ coordinate system
\begin{equation}
    \left[-\frac{\hbar^2}{2m}\nabla^2 + V(\bar{x})\right] \psi(\bar{x},\bar{y}) = E \psi(\bar{x},\bar{y}).
\end{equation}
The incident momenta components are $k_{\bar{x}}=\sqrt{\frac{2mE}{\hbar^2}}\cos{\theta}$ and $k_{\bar{y}}=\sqrt{\frac{2mE}{\hbar^2}}\sin{\theta}$, where $E$ is the kinetic energy in region I (with zero potential). Due to the translational invariance along $\bar{Y}$, $k_{\bar{y}}$ is conserved, and we employ the separable ansatz, $\psi(\bar{x},\bar{y})=\Phi(\bar{x}) \chi(\bar{y})=\Phi(\bar{x})e^{i k_{\bar{y}}\bar{y}}$. The potential step exists at $\bar{x}=0$: $V(\bar{x})=0$ for $\bar{x}<0$ (region I) and $V(\bar{x})=V_0$ for $\bar{x}>0$ (region II). To proceed, we solve for the wavefunction separately in each region, which are given as, $ \Phi_1(\bar{x})=e^{i k_{\bar{x}} \bar{x}} +r\:e^{-i k_{\bar{x}} \bar{x}},\:
\Phi_2(\bar{x})=t\:e^{i k_{\bar{x}}' \bar{x}},$ where, $r$ and $t$ denote the reflection and transmission coefficients, respectively, while $k'$ represents the magnitude of the wavevector in the region with potential, defined as, $k'=\sqrt{\frac{2m(E-V_0)}{\hbar^2}}$. Specifically, translational invariance along $\bar{Y}$ implies $k'_{\bar{y}}=k_{\bar{y}}$. Next, we evaluate the probability current, $J_{\alpha}=-\frac{i\hbar}{2m}(\psi^*\frac{\partial \psi}{\partial \alpha}-\psi\frac{\partial \psi^*}{\partial \alpha})$, in direction $\alpha$. Applying this to region II, we obtain
$        J_{\bar{X}}=\frac{\hbar k'_{\bar{x}}}{m}|t|^2, \quad J_{\bar{Y}}=\frac{\hbar k_{\bar{y}}}{m}|t|^2 .$

To estimate the transverse response, $J_A$, we calculate the current density along the $Y$ direction. The component of the current transverse to the sample can then be obtained, through a straightforward coordinate transformation, as
\begin{equation}
\begin{split}
   J_A&=-J_{\bar{Y}}\cos{\theta}+J_{\bar{X}}\sin{\theta} \\
   &= \frac{\hbar}{m} |t|^2 (-k_{\bar{y}}\cos{\theta}+ k'_{\bar{x}}\sin{\theta}) \\
   &= \frac{\hbar}{m} |t|^2 (-k \sin{\theta}\cos{\theta}+k'\cos{\theta'}\sin{\theta}).
\end{split}
\end{equation}
By applying Snell's law of refraction, and substituting the expressions for $\theta'$ and $k'$ we obtain
\begin{equation}
    J_A=|t|^2 \sqrt{\frac{2E}{m}}\sin{\theta}\: \left(-\cos{\theta}+\sqrt{\cos^2{\theta}-\frac{V_0}{E}}\:\right).
\end{equation}

Imposing the boundary conditions on $\psi$ and its derivative $\frac{\partial \psi}{\partial \bar{x}}$ at $\bar{x}=0$, the transmission probability $|t|^2$ is determined as $ t=\frac{2 k_{\bar{x}}}{k_{\bar{x}}+k'_{\bar{x}}}= 2\:(\:1+\sqrt{1-\frac{V_0}{E\cos{\theta}}}\:)^{-1}$. Using these results, we arrive at the final expression for the Hall-like current in the single-electron case,
\begin{equation}
    J_A=\sqrt{\frac{2E}{m}}  \frac{4 \sin{\theta}}{\abs{1+\sqrt{1-\frac{V_0}{E\cos{\theta}}}}^2} \left(\sqrt{\cos^2\theta-\frac{V_0}{E}}-\cos{\theta}\right). \label{ja}
\end{equation}

This expression satisfies two consistency checks.
First, at zero tilt angle of the potential, the transverse response must vanish; this is satisfied as $J_A(\theta=0)=0$. Second, in the absence of a potential barrier, the response should vanish; this is also satisfied as $J_A(V_0=0)=0$.\\

\begin{figure}[t]
    \centering
    \includegraphics[width=0.35\textwidth]{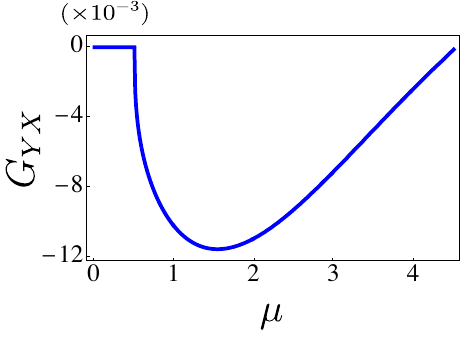}
    \caption{\label{Fig:Theory} \textbf{Refraction-driven Hall-like conductance from analytical calculations.} The calculated transverse conductance, $G_{YX}$, is shown as a function of the Fermi energy $\mu$. The parameters are fixed at a barrier strength $V_{0} = 0.5$ and a tilt angle $\theta = 0.025 ~\pi$. A finite Hall-like response emerges when the Fermi energy exceeds the barrier height $V_{0}$. The conductance is expressed in units of $e^{2}/h$.}
\end{figure}


\begin{figure*}[t]
    \centering
    \includegraphics[width=0.95\textwidth]{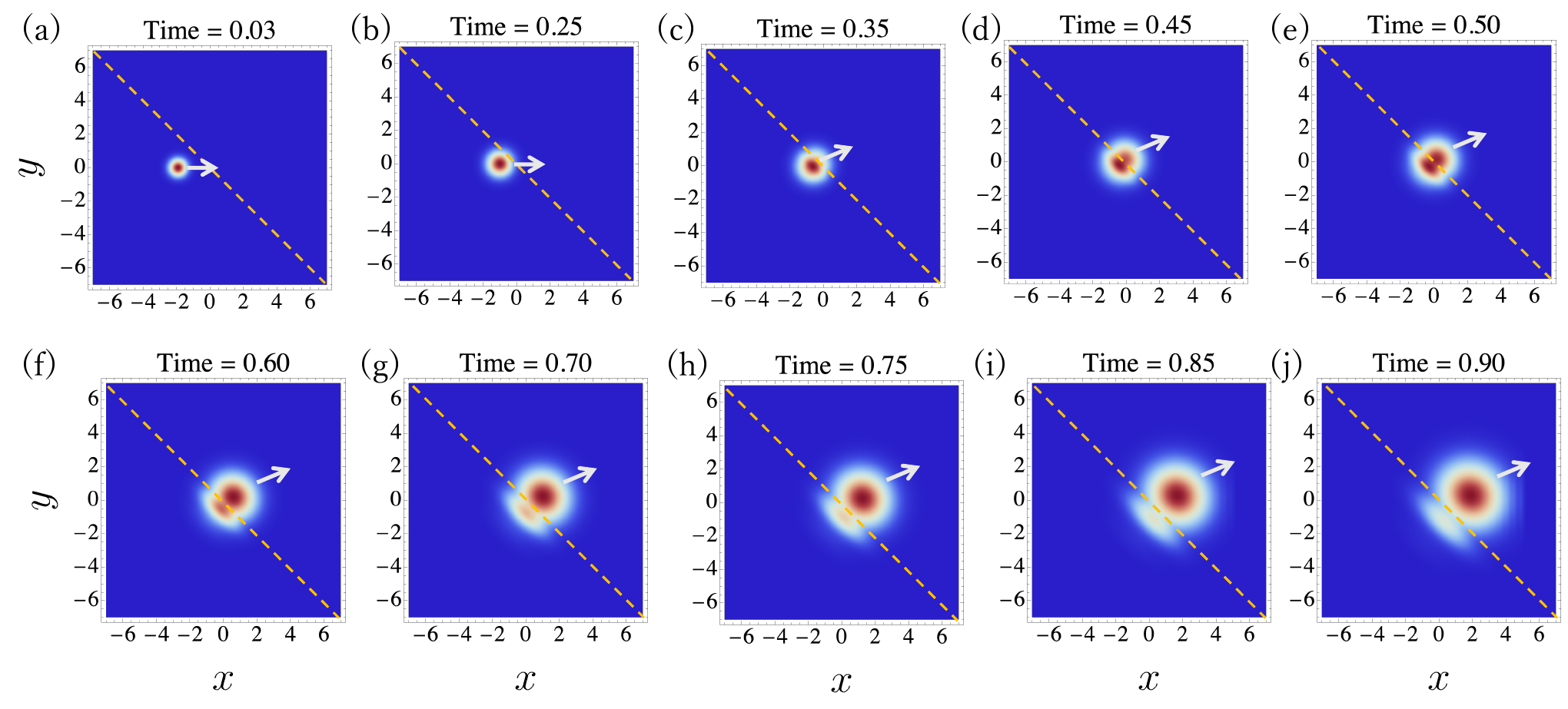}
    \caption{\label{Fig:TimeEvolution}\textbf{Time dynamics illustrating the refraction-driven Hall-like effect.} The time evolution of a Gaussian wave packet initially localized at $(-2,0)$ is shown. The packet is assigned an initial momentum along the $x$-direction, resulting in horizontal propagation prior to interaction with the tilted potential barrier. Upon transmission through the barrier, the wave packet acquires a finite positive momentum component along the $y$-direction, thereby demonstrating the transverse transport of electrons. The dynamics is computed by numerically solving the time-dependent Schr\"odinger equation. The dashed line indicates the interface of the potential barrier. The potential takes the value $V_{0}$ above the line and vanishes below it. In this case, the barrier is tilted at an angle $\theta = \pi/4$ with respect to the axes, and the barrier strength is $V_{0}=-0.5$.}
\end{figure*}

\begin{figure*}[t]
    \centering
    \includegraphics[width=0.95\textwidth]{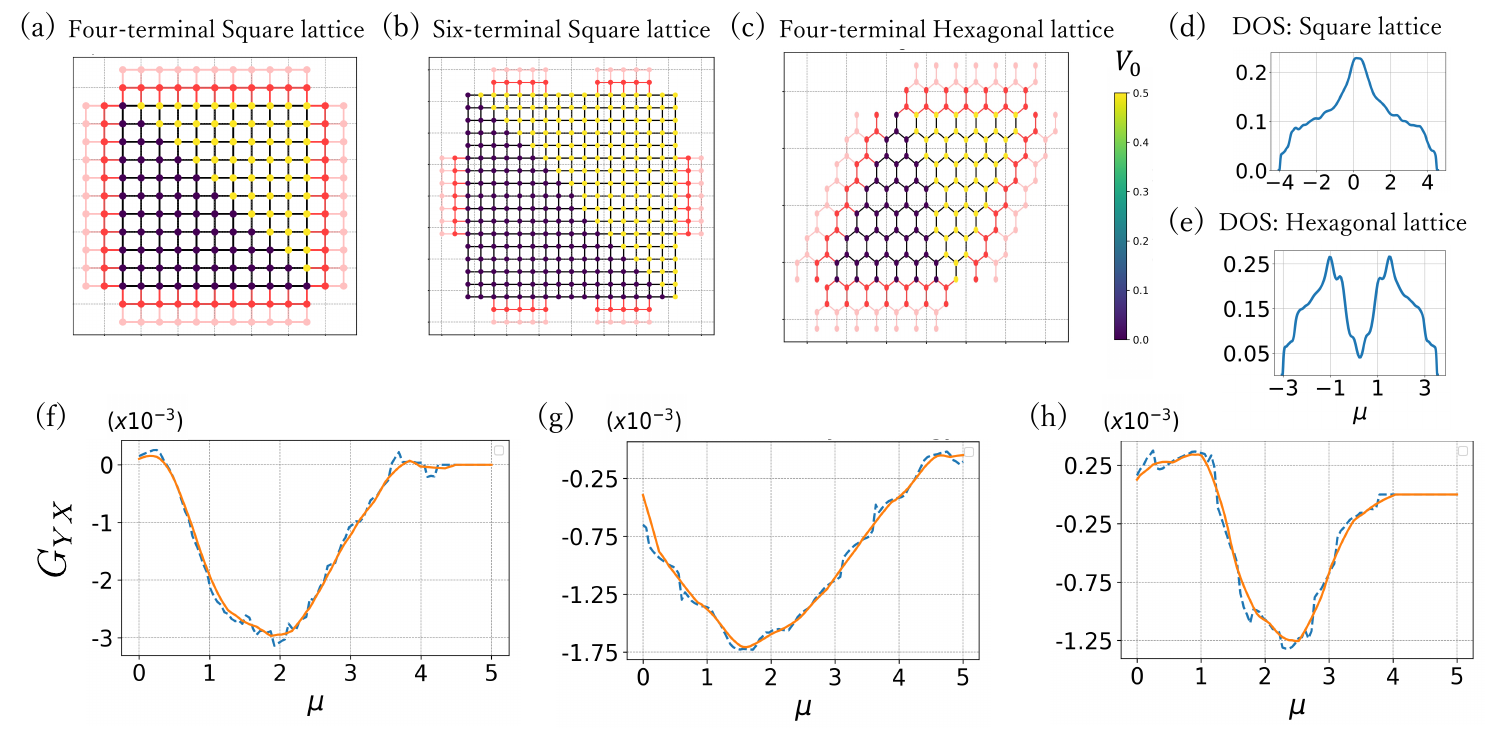}
    \caption{\label{Fig:Kwant}\textbf{Refraction-driven transverse conductance from lattice simulations.} The upper panels (a)-(c) depict the schematic structure of the simulated device geometries, while the corresponding Hall-like conductance are shown in the lower panels (f)-(h). Panels (a) and (f) present results for a four-terminal square lattice; panels (b) and (g) correspond to a six-terminal square lattice; and panels (c) and (h) show a four-terminal hexagonal lattice. In all cases, the potential barrier has a tilt angle of $\theta = 0.025~\pi$ and strength $V_{0} = 0.5$. A longitudinal current is applied between the horizontal leads, and the transverse voltage is measured across the vertical leads. The numerical simulations reproduce the analytical trends within the regime of interest. Blue dashed lines denote the moving average of the raw data, while orange curves denote the smoothened profiles. Panels (d) and (e) show the normalized density of states (DOS) with the chemical potential, $\mu$, for the square and hexagonal lattice structures, respectively. All conductance values are expressed in units of $e^{2}/h$. Here the lattice dimensions are $L_X=20, L_Y=200$ for square lattices and $L_X=20, L_Y=400$ for hexagonal lattice geometries.}
\end{figure*}

The more physically relevant situation corresponds to electrons incident at all possible angles, rather than an electron incident at a specific angle. We
therefore consider the system depicted in Fig.~\ref{Fig:Schematic}(b), which is finite along $X$ and is assumed to be infinite along $Y$
(hence there is translational symmetry along
$\bar{Y}$). The sample is connected to a source and a drain along $X$, thus applying an external bias. 
Electrons may impinge 
from the source on to the potential barrier at a continuum of angles, and their collective contributions determine the net Hall-like response, if any. To analyze this, we recall the coordinate transformation to the device frame,
$\hat{X}=\hat{\bar{X}}\cos{\theta}-\hat{\bar{Y}}\sin{\theta}$, $\hat{Y}=\hat{\bar{X}}\sin{\theta}+\hat{\bar{Y}}\cos{\theta}$, and we focus on the response along $\hat{Y}$. For a single electron transmitting into the potential barrier, the current density is
\begin{equation}
    \delta I_Y= |t|^2 \frac{\hbar^2}{m} (k'_{\bar{x}}\sin{\theta} +k_{\bar{y}} \cos{\theta}).
\end{equation}
Now, electrons may be incident at an angle
$\phi$, such that $k_{\bar{x}}=k\cos({\phi-\theta})$ and $k_{\bar{y}}=k\sin({\phi-\theta})$, where $\phi \in [-\frac{\pi}{2},\frac{\pi}{2}]$. The source is assumed to supply a chemical potential $\mu_S=\mu +\delta \mu$, while the drain is held at $\mu_D=\mu$. Consequently, the electronic energies are constrained by the chemical potentials to lie in the
range $\mu < \frac{\hbar^2 k^2}{2m} < \mu+\delta \mu$. The conservation of energy implies that $k'_{\bar{x}}=\sqrt{k_{\bar{x}}^2-2mV_0/\hbar^2}$. Combining this with the expression given above for the transmission probability $|t|^2$, we obtain the total transverse current
\begin{equation}
\begin{split}
    \delta I_Y= \frac{\hbar^2}{m \pi^2} \int dk_{\bar{x}}\: dk_{\bar{y}} \: \frac{k_{\bar{x}}^2}{k_{\bar{x}}+\sqrt{k_{\bar{x}}^2-2mV_0/\hbar^2}}\: ~\times \\ 
    \left(\sqrt{k_{\bar{x}}^2-2mV_0/\hbar^2}\: \sin{\theta}+k_{\bar{y}}\:\cos{\theta} \right).
\end{split}
\end{equation}

We now transform to radial coordinates, $\int dk_{\bar{x}}\: dk_{\bar{y}} \rightarrow \int k\: dk \int d\phi$. Assuming the chemical potential bias, $\delta\mu$, is small, we make the approximation, $k^2 \simeq 2m\mu/\hbar^2$. This simplifies the integral
as $\int k \:dk = \frac{1}{2}\int d (k^2) \simeq \frac{m \delta \mu}{\hbar^2} $. We then obtain the differential transverse conductance, $G_{YX}\:=\delta I_Y/\delta\mu$,

\begin{widetext}
\begin{equation}
    G_{YX}\:(\mu,V_0,\theta)=\frac{2 m \mu}{\pi^2 \hbar^2}\:\int d\phi\: \frac{\cos^2(\phi-\theta)}{(\:\cos{(\phi-\theta)}+\sqrt{\cos^2(\phi-\theta)-\nu^2}\:)^2}
    \left(\sqrt{\cos^2(\phi-\theta)-\nu^2}\:\sin{\theta}+\sin{(\phi-\theta)}\cos{\theta} \right), \label{G}
\end{equation}
\end{widetext}
where $\theta$ is the tilt angle of the potential, and $\nu^2= V_0^2/\mu^2$. We can again
verify two limiting cases from Eq.~\eqref{G}. First, in the absence of a potential, $V_0=0$, we find $G_{YX}\sim \int d\phi [\cos{(\phi-\theta)}  \sin \theta + \sin{(\phi-\theta)} \cos \theta] = 0$.
Second, for a potential that is not tilted, $\theta=0$, we obtain $G_{YX}\sim \int d\phi \frac{\cos^2\phi \sin \phi}{(\cos \phi +\sqrt{\cos^2 \phi -\nu^2})^2}=0$, since the integrand is an odd function of $\phi$. We note 
here that our system does not have rotational symmetry; as a result, the conductance matrix $G_{ij}$ is neither symmetric nor antisymmetric in general. The symmetries
and relations between the different matrix elements
of $G_{ij}$ are discussed in Appendix A.

The transverse response, as derived from our analytical calculations, is presented in Fig.~\ref{Fig:Theory}. The Hall-like conductance is shown as a function of $\mu$. When $\mu<V_0$ , no response is observed, as the electrons undergo reflection and are unable to penetrate the potential barrier. However, for $\mu>V_0$, a non-zero $G_{YX}$ is obtained, leading to a finite transverse response in the system, even in the absence of a magnetic field. \\

\section{Dynamics under the refraction-driven Hall-like transport}
To illustrate the refraction-driven transverse effect, we next consider the dynamical evolution of a Gaussian wavepacket in a typical refraction setup. We construct a square lattice and engineer a tilted negative potential at an angle \(\theta = \pi/4\) for visual clarity. A Gaussian wave packet, initially centered at \((-2, 0)\) within the system, is introduced and its time evolution is tracked using the time-dependent Schr\"odinger equation. The wave packet is initially given a momentum along the \(x\)-direction, with no component of momentum in the transverse \(y\)-direction. (The details of the
calculations of the dynamics are given in Appendix B).
As depicted in Fig.~\ref{Fig:TimeEvolution}, the wave packet initially propagates horizontally towards the potential barrier. In addition to its motion along the \(x\)-direction, the wave packet exhibits a spreading, typical of Gaussian time-evolution. Importantly, when the wave packet encounters the tilted potential barrier, shown by the dashed line, a distinct feature emerges -- the wave packet acquires a component of momentum in the transverse \(y\)-direction, indicating a Hall-like response. This transverse component is associated with the portion that is refracted into the region with the potential.
Since our simulation uses a negative potential, the electron wave packet refracts towards the normal.

These observations are in direct agreement with both our analytical calculations and the results obtained from the numerical simulations (detailed in the next section). The emergence of a transverse momentum component from an initially purely longitudinal wave packet provides evidence of the Hall-like response in this system, driven by refraction at the tilted potential barrier. The dynamics observed confirm the validity of our analytical framework and illustrate how such a transverse effect can arise due to the interaction of the wave packet with a tilted potential, \emph{without} the need for a magnetic field.\\

\section{Refraction-induced transverse conductance for lattice models}
To further complement our analytical results, we next simulate a refraction geometry for lattice models. We examine several device geometries of direct experimental feasibility using the Kwant simulation package~\cite{groth2014kwant}, and compute the Hall-like response by employing scattering matrix calculations. The different geometries are schematically illustrated in the upper panels of Fig.~\ref{Fig:Kwant} (a)-(c). We consider square lattice geometries with four- and six-terminals, along with the hexagonal lattice geometry. For each configuration, we model the scattering region by a nearest-neighbor tight-binding model. (The details
of the lattice model are presented in Appendix C).
The tilted potential is incorporated within the scattering region, with a constant value \(V_0 = 0.5\), chosen to closely mirror the conditions of our analytical calculations. The leads, shown in red in Fig.~\ref{Fig:Kwant}, are attached to the scattering region and extend to infinity. The input current is constrained to flow from the left to the right terminals in all cases, while the transverse voltage is measured between the upper and lower terminals, enabling the calculation of the Hall-like conductance, $G_{YX}$. This is shown in the lower panels of Fig.~\ref{Fig:Kwant} (f)-(h). The conductance results, represented by blue dashed lines in these panels, are computed using a moving average, while the orange curves are the smoothed profiles to highlight the general trend.
In all of the simulated geometries, a non-zero transverse response is observed as a function of the Fermi energy, \(\mu\), and the magnitude of this response is of the same order as predicted by our analytical calculations. While an exact match between the analytical and simulated results is not expected, the general agreement establishes the validity of our proposed effect. Our theory assumes a quadratic-dispersion electron model for simplicity, whereas the numerical simulations use a more realistic tight-binding lattice model, which accounts for additional effects such as the band structure. A noticeable deviation from the theoretical predictions arises when the transverse conductance becomes zero after a certain value of \(\mu\). This is due to the density of states being non-zero only in a finite energy window, as illustrated in Fig.~\ref{Fig:Kwant} (d)-(e), for both square and hexagonal lattice geometries. When \(\mu\) lies outside this energy window, no states are available for conduction, causing the Hall-like response to vanish. However, within this energy window, a non-zero transverse response is clearly observed, confirming the viability of the proposed geometries for experimental realization. Overall, these simulations corroborate our analytical predictions and suggest that the refraction-driven transverse response is a robust feature, observed across different lattice models. Finally, we note that our analysis assumes ballistic transport. Graphene, with its hallmark hexagonal lattice, is well-known to exhibit ballistic transport features~\cite{du2008approaching,mayorov2011micrometer,baringhaus2014exceptional,chen2016electron}, making it a promising platform for the experimental realization of our predicted refraction-induced Hall-like effect.\\

\section{Summary and outlook}
In summary, we proposed and developed a new Hall-like response that arises without an external magnetic field or breaking of time-reversal symmetry. We demonstrated that a tilted potential interface induces a refraction-like behavior of electron wave packets, producing a finite transverse current and a Hall-like conductance. This transverse response emerges purely from the geometric features of the potential landscape and represents a fundamentally distinct mechanism for generating transverse transport in time-reversal-invariant systems. Our analytical framework, supported by numerical simulations on different lattice models and device geometries, consistently captures the essential features of the effect. The real-time wave packet dynamics further provides a direct visualization of this effect, revealing its geometric origin and a robustness of the transverse response. Our findings demonstrate that geometric control of potential barriers offers a powerful route to manipulate Hall-like transport in mesoscopic systems.
Tilted potential interfaces may occur naturally in certain experimental settings or can be engineered by substrates or structural asymmetries that induce spatially varying onsite potentials~\cite{muggli2001refraction,repp2004snell,cheianov2007focusing,low2009electronic,moghaddam2010graphene,gao2011giant,lee2015negative_refraction,baeumer2015ferroelectrically,li2016gate,liu2017directional_beams_graphene,ye2023reconfigurable}. Overall, our results open as-yet-unexplored avenues for designing devices that exploit geometric effects to harness unconventional Hall responses.\\

\section*{Acknowledgments}
R.S. is supported by the Prime Minister's Research Fellowship (PMRF). R.S. thanks S. Mukerjee for useful discussions. A.N. acknowledges DST CRG grant (CRG/2023/000114) for support. D.S. thanks SERB, India for support through the project 
JBR/2020/000043.

\appendix

\section{Symmetries and relations between the conductance matrix elements}

Our setup preserves time-reversal symmetry. However, due to the tilted potential interface the rotational symmetry is broken, which would otherwise relate transport along the $x$ and $y$ directions in a simple way. As a result, the off-diagonal conductance values are not expected to be symmetric/antisymmetric to each other, i.e., $G_{XY}\neq \pm \:G_{YX}$ in general. 
To understand more precisely why there is no
symmetry or antisymmetry, we first describe the
geometry of our system and our conventions.
We consider a system in which the voltage and
current are measured in the $x$ and $y$ directions as usual. Next, there is an
interface whose normal points in a direction
which makes an angle $\theta$ (taken to be positive in
the counterclockwise direction) with respect to
the $+ {\hat x}$ direction. The Hall-like conductances
are then denoted as $G_{YX} (\theta)$ and
$G_{XY} (\theta)$, which denote the current
in the ${\hat y} ~({\hat x})$ direction when a potential gradient is applied in the ${\hat x} ~
({\hat y})$ direction respectively. Upon noting
that an wave incident from the region with $V=0$ bends away from the normal
when it enters the region with $V=V_0$, we find
from simple geometric considerations that
$G_{YX} (\theta) = G_{XY} (\pi/2 - \theta)$. (In deriving this we have used the facts
that $G_{YX} (-\theta) = - G_{YX} (\theta)$ and $G_{-X,Y} (\theta) = - G_{XY} (\theta)$). This relation shows that
it is sufficient to calculate $G_{YX} (\theta)$ for different values of $\theta$; 
we do not then need to calculate 
$G_{XY} (\theta)$ separately. We conclude
that there is no simple relation between 
$G_{YX} (\theta)$ and $G_{XY} (\theta)$
in general. However, for the special case
$\theta = \pi/4$, we see that there is a 
symmetry $G_{YX} (\pi/4) = G_{XY} (\pi/4)$.

Due to the broken rotational symmetry, a simple symmetric/anti-symmetric relation between the off-diagonal conductance components does not hold. Our setup provides a purely geometric origin of the transverse conductance. There are several examples of Hall effects preserving time-reversal symmetry, for instance, the spin Hall insulator~\cite{murakami2004}, the Magnus Hall effect~\cite{costa2018gravitational,papaj2019magnus,mandal2020magnus}, non-linear Hall effects~\cite{sodemann2015quantum,du2021nonlinear,ortix2021nonlinear,bandyopadhyay2024non}, the valley Hall effect~\cite{xiao2007valley,mak2014valley,gorbachev2014detecting}, and the anisotropic anomalous Hall effect~\cite{koizumi2023}. Our proposed refraction-driven Hall-like effect naturally falls within this broader class of time-reversal–symmetric transverse responses.\\

\section{Calculations for the dynamics of a Gaussian wave packet under refraction}

We simulate the time evolution of a two-dimensional Gaussian wave packet incident on a tilted potential barrier. We initialize the state with the following form,
\begin{equation}
    \psi_0(x,y)\sim e^{-\frac{(x-x_0)^2+(y-y_0)^2}{2 \sigma^2}} e^{i(k_x x+k_y y)}.
\end{equation}
Here $(x_0,y_0)$ is the initial wave packet center which we set to $(-2,0)$, and $\sigma$ is the width of the wave packet controlling how spread out the Gaussian is. Initially $\sigma=0.5$, but it then increases under time evolution. $k_x$ and $k_y$ are the initial momentum components, and we have chosen $k_x=1$ and $k_y=0$. Further, the Gaussian wave packet is normalized so that the total probability density integrates to 1 over the $(x,y)$ domain.\\
Using this initialization, we solve the time-dependent Schr\"odinger equation in two dimensions to obtain the dynamics of the wave packet. The Schr\"odinger equation is given as follows:
\begin{equation}
    i\hbar \frac{\partial\psi}{\partial t}= -\frac{\hbar^2}{2m} \nabla^2 \psi + V(x,y) \psi,
\end{equation}
where, $V(x,y)$ is the potential which includes the tilted barrier. We consider a discretized two-dimensional lattice of size $8 \times 8$, with 400 grid points along each spatial direction. The system is evolved up to a total time $t_{\textup{max}} = 1$ using a time step of $dt = 0.001$. At each time step, the dynamics are computed by solving the discretized form of the time-dependent Schr\"odinger equation on the lattice. The probability density $|\psi(x,y,t)|^2$
is then plotted in the $x-y$ plane at successive times; the snapshots shown in Fig.~\ref{Fig:TimeEvolution} correspond to these time slices and visualize the 
deflection/refraction of the wave packet upon encountering the tilted interface.\\

\section{Details of the lattice model calculations}

In Fig.~\ref{Fig:Kwant}(a,b), the scattering geometry is a two-dimensional square lattice with a tilted potential barrier. The tight-binding model for these geometries is given by
\begin{equation}
    H=t\sum_{\langle i,j \rangle} (c^\dagger_i c_j + c^\dagger_j c_i) +\sum_i V(r_i) c^\dagger_i c_i,
\end{equation}
where, $\langle i,j \rangle$ denotes all pairs of nearest neighbors. $V(\vec{r})$ is the onsite potential due to the barrier which takes the value $V_0=0.5$ when $x>-\tan(\theta)\:y$ and is $0$ otherwise. The hopping amplitude $t=1$ sets the energy scale of the system and the tilt angle $\theta=45^o$.

We simulate the electron transport in the two-dimensional lattice using {\sc kwant}, a Python package for quantum transport simulations~\cite{groth2014kwant}. We specifically compute the transverse conductance for the sample by connecting semi-infinite leads to the scattering region, as presented in Fig.~\ref{Fig:Kwant}. The hopping between neighboring sites in the leads is also set to match the hopping in the tight-binding lattice $t_\textup{lead}=t=1$, ensuring mode matching at the lead interfaces. We specify the chemical potential range for which we want to calculate the transverse response. For each energy, the scattering matrix is calculated which provides information about the transmission and reflection properties of the system. We impose the condition that current ($I$) flows only from the left to the right lead in each case. Then, {\sc kwant} computes the transmission matrix, to calculate the conductance between the specified vertical leads. The horizontal leads thus act as current injectors and absorbers while the vertical leads act as voltage probes. The Hall-like voltage, $V_\textup{Hall}$ is defined as the potential difference between the vertical leads. From this, we can compute the resistance, $R_H=V_\textup{Hall}/I$, and the corresponding transverse conductance, $G_{YX}=1/R_H$.

For the geometry in Fig.~\ref{Fig:Kwant}(c), we repeat all the same calculations with the hexagonal (honeycomb) lattice system. Here, the tight binding model is as follows,
\begin{eqnarray}
    H &=&  t \sum_{\langle i,j \rangle} (c^\dagger_{i,A}c_{j,B}+ c^\dagger_{j,B}c_{i,A})\nonumber \\
    && +\sum_i V(r_i) (c^\dagger_{i,A}c_{i,A}+ c^\dagger_{i,B}c_{i,B}),
\end{eqnarray}
where $A,B$ label the two sublattices of the honeycomb lattice. $V(\vec{r})=V_0$ for $x>-\tan(\theta)\: y$ and $0$ otherwise. We fix $V_0=0.5$ and $\theta=45^o$ for our simulations.

\bibliography{Ref}

\end{document}